\documentclass[runningheads,a4paper]{llncs}
\usepackage{graphicx}
\usepackage{float}
\usepackage{multirow}
\usepackage{wrapfig}

\graphicspath{{img/}}

\newcommand{\keywords}[1]{\par\addvspace\baselineskip
\noindent\keywordname\enspace\ignorespaces#1}

\begin{document} 
\mainmatter

\title{Change Patterns for Model Creation: Investigating the Role of Nesting
Depth\thanks{This research is supported by Austrian Science Fund (FWF):
P23699-N23. The final publication is available at Springer via
http://dx.doi.org/10.1007/978-3-642-38490-5\_19}\\\vspace{0.4cm}\small{Position
Paper}\vspace{-0.3cm}}
\titlerunning{Change Patterns for Model Creation} 
\author{Barbara Weber\inst{1} \and Jakob Pinggera\inst{1} \and Victoria
Torres\inst{2} \and Manfred Reichert\inst{3}}
\authorrunning{Weber et al.}
 
\institute{University of Innsbruck, Austria\\
\email{{barbara.weber, jakob.pinggera}@uibk.ac.at}  
\and Universitat Polit\`ecnica de Val\`encia, Spain\\
\email{vtorres@pros.upv.es}
\and University of Ulm, Germany\\
\email{manfred.reichert@uni-ulm.de}
}   
    
\maketitle 

\vspace{-0.7cm}
\begin{abstract}
Process model quality has been an area of considerable research efforts. In this
context, the correctness-by-construction principle of change patterns offers a
promising perspective. However, using change patterns for model creation imposes
a more structured way of modeling. While the process of process modeling (PPM)
based on change primitives has been investigated, little is known about this
process based on change patterns and factors that impact the cognitive
complexity of pattern usage. Insights from the field of cognitive psychology as
well as observations from a pilot study suggest that the nesting depth of
the model to be created has a significant impact on cognitive complexity. This
paper proposes a research design to test the impact of nesting depth on the
cognitive complexity of change pattern usage in an experiment.
\vspace{-0.3cm}
\end{abstract}

\vspace{-1.1cm}
\section{Introduction}
\vspace{-0.1cm}
Much conceptual, analytical, and empirical research has been conducted during the
last decades to enhance our understanding of conceptual modeling. In particular,
process models have gained significant importance due to their fundamental role
for process-aware information systems. Even though it is well known
that a good understanding of a process model has a direct and measurable impact
on the success of any modeling initiative \cite{DBLP:journals/dss/KockVDD09},
process models display a wide range of quality problems impeding their
comprehensibility and maintainability \cite{DBLP:journals/dke/MendlingVDAN08}. 

To improve process model quality, change patterns offer a promising
perspective~\cite{WRR08}. Instead of creating a process model using
change primitives (e.g., add node, add edge) high-level change operations
combining several change primitives are used as basic building blocks for model
creation. Examples of change patterns include the insertion and deletion of
process fragments or their embedding in loops. Particularly appealing is
correctness-by-construction~\cite{WRR08}, i.e., the modeling environment 
provides only patterns to the process modelers, which ensure that a sound
process model is transformed into another sound model.

The use of change patterns implies a different way of creating process models,
since correctness-by-construction imposes a structured way of modeling by
enforcing block structuredness. Irrespective of whether change patterns or change
primitives are used, model creation requires process modelers to construct a
mental model (i.e., \textit{internal representation}) of the requirements to be
captured in the process model~\cite{Sof+12}. In a subsequent step, the mental
model is mapped to the constructs provided by the modeling language creating an
\textit{external representation} of the domain~\cite{Sof+12}. While the
construction of the mental model presumably remains unaffected, the use of change
patterns leads to different challenges concerning pattern selection and
combination for creating the external representation. In particular, process
modelers might have to look several steps ahead to construct a certain process
fragment, which constitutes a major difference compared to the use of change
primitives, which do not impose any structural restrictions.
  
The process of creating process models based on change primitives has caused
significant attention leading to a stream of research on the \textit{process of
process modeling} (PPM) \cite{Sof+12,PZW+12,CVR+12,PSZ+12}. This research is
characterized by its focus on the formalization phase of model creation, i.e.,
the modeler's interactions with the modeling environment~\cite{PZW+12}. Still,
little is known about the PPM when utilizing change patterns. To fill this gap,
we conducted a pilot study with 16 process modelers \cite{WPT+XX}, which
indicated that the cognitive complexity imposed by change pattern usage is highly
related to the structure of the process model to be created, in particular the
nesting depth of the model. Modelers did not face any major
problems when constructing simple process fragments, e.g., when inserting
activities in sequences, making an activity optional, or inserting an activity in
parallel. Faced with more complex control flow structures, in turn, the
structural restrictions imposed by change patterns led to considerable problems
(i.e., detours or incorrect models). These observations were underlined by
feedback of the participants who appreciate the correctness-by-construction
guarantees, but feel restricted when faced with complex control flow constructs.
To further investigate these observations this paper proposes a research design
to test the influence of nesting depth on the cognitive complexity of change
pattern use.

\vspace{-0.55cm}
\section{Cognitive Foundations of Problem Solving}\label{cognFound}
\vspace{-0.25cm}

We consider the creation of process models to be a complex problem solving task.
Problem solving has been an area of vivid research in cognitive
psychology for decades. Therefore, we turn to cognitive psychology to understand
the processes followed by humans when solving a problem like creating a process model.

\textbf{Schemata.} The human
brain contains specialized regions contributing different functionality to the
process of \emph{solving complex problems}. \emph{Long-term memory} is
responsible for permanently storing information and has an essentially unlimited
capacity, while in \emph{working memory} comparing, computing and reasoning take
place~\cite{Gray07}. Although the latter is the main working area of the brain,
it can store only a limited amount of information, which is forgotten after
20--30 seconds if not refreshed~\cite{Trac79}. The question arises how
information can be processed with such limited capacity. The human mind organizes
information in interconnected \emph{schemata} rather than in
isolation~\cite{Gray07}. Those schemata, stored in long-term memory, incorporate
general concepts of similar situations~\cite{Gray07}. Whenever situations similar
to a schema arise, the latter is retrieved to help organizing information by
creating \emph{chunks} of information that can be processed
efficiently~\cite{JTPA81}. To illustrate how chunking actually influences the
understandability of process models consider a fragment with one alternative
branch. Unexperienced modelers may use three chunks to store such process: one
for each XOR- gateway and one for the activity. In contrast, an expert may
recognize the pattern for optional  activities, i.e., a schema for optional
activities is present in long-time memory, allowing the storage of the
entire process fragment in one working memory slot.

  
\textbf{Problem-Solving Strategies.} 
Novices confronted with an unfamiliar problem cannot rely on
specialized problem solving strategies. Instead, an initial skeletal
\emph{plan} is formed~\cite{Rist89}. Then, they utilize general problem solving
strategies, like means-ends analysis, due to the lack of more specific strategies for the task~\cite{KaNe84}.
Means-ends analysis can be described as the continual comparison of the problem's current
state with the desired end product. Based on this, the next steps are selected until a
satisfying solution is found~\cite{KaNe84}. After applying the constructed plan,
it can be stored in long-term memory as \emph{plan schema}~\cite{Rist89}. For
this, task-specific details are removed from the plan schema resulting in a plan
schema that can be automatically applied in similar situations~\cite{Ande82}.
When confronted with a problem solving task in the future, the appropriate plan
schema is selected using case-based reasoning~\cite{GuCu88}. The
retrieved plan schema provides the user with structured knowledge
that drives the process of solving the problem~\cite{JTPA81,GuCu88}. Plan
schemata allow experts to decide what steps to apply to end up with
the desired solution~\cite{Swel88}. If the plan schema is well developed, an
expert never reaches a dead end when solving the problem~\cite{Broo77}.

Plan schemata seem important when creating process models based on change
patterns since patterns cannot be combined in an arbitrary manner,
especially when complex control-flow structures have to be created. If no plan
schema is available on how to combine patterns to create the desired
process model, modelers have to utilize means-ends analysis until a satisfying
solution is found. This behavior is more likely to result in detours and
decreased modeling speed. Moreover, the cognitive complexity for conducting
means-ends analysis increases when confronted with complex control-flow structures like deeply nested blocks, making it more difficult to reach the correct solution. In addition, respective
structures require the modeler to possess schemata to process larger chunks
of information beforehand (i.e., increased need for look-ahead) to be able to model the
respective fragment fast and without any detours. As a
consequence, mental effort increases as well as the probability for detours. 

\vspace{-0.4cm}

\vspace{-0.05cm}
\section{Research Design}\label{research Model}

\vspace{-0.15cm}

Based on the cognitive foundations in Sect. \ref{cognFound} we
propose a research design to investigate the influence of nesting depth on the
cognitive complexity of change patterns usage for model creation
by means of a controlled experiment.

\textbf{Subjects.} As explained in Sect. \ref{cognFound}, novices and experts
differ in their problem solving strategies. While novices have to rely on
general problem solving strategies like means-ends analysis, experts can rely on
plan schemata. Since process modelers in practical settings are often not expert
modelers, but rather casual modelers with a basic amount of
training~\cite{PZW+10}, we do not require modeling experts for our study. To
avoid, however, that difficulties are caused by unfamiliarity with the tool,
rather than by difficulties with the tasks themselves, we require subjects to be
moderately familiar with process modeling as well as change patterns. For this,
subjects are trained using theoretical
backgrounds of change patterns, but also obtain hands-on experience in the
creation of process models using change patterns to guarantee that they are
sufficiently literate in change pattern usage. Regarding
process modeling experience and experience in change patterns usage we assume a
relatively homogeneous group, which is tested ex-post. Choosing subjects
moderately familiar with process modeling and change patterns usage allows
us to make statements about casual modelers that cannot be generalized to
modeling experts.
  
\textbf{Independent Variable and Factor
Levels. } As independent variable we consider the \textit{nesting depth} of the
solution model with factor levels: \textit{high} and \textit{low}.

\textbf{Objects.} As outlined in Sect. 1, the creation of a process model
requires the process modelers to construct a mental model (i.e., \textit{internal
representation}) of the requirements to be captured in the process
model~\cite{Sof+12} and to map this mental model to the constructs provided by
the modeling language (i.e., creating an \textit{external representation} of the
domain~\cite{Sof+12}). Tasks should be designed such that it can be ensured that
observed difficulties are caused by change patterns usage rather than problems
understanding the domain and constructing the mental model (which would also
exist when using change primitives for model creation). Therefore,
to factor domain influences out, participating subjects are asked to re-model a
process (denoted as reference model in the following) starting from an empty
modeling canvas. For this, process designers have to apply a sequence
of change patterns to incrementally re-construct the given reference model
starting from the empty model canvas.
In addition, activities are labeled A, B, C to reduce the potential impact
of domain knowledge.

Since the experiment aims to compare the cognitive complexity of using change
patterns depending on the nesting depth, two versions (with high and low nesting
dept) of the modeling task have to be designed, making sure that both tasks
differ only in their nesting depth and not in other model characteristics. From
research into process model quality we know, for example, that the size of the
process model impacts model comprehension \cite{DBLP:journals/dss/MendlingSR12}
and that different control-flow constructs do not have the same cognitive
complexity \cite{DBLP:conf/caise/FiglL11}. As a consequence, these factors have
to be controlled when designing the material for the experiment. Our intention
therefore is to choose 2 models with the same number of activities, the same
change patterns, and the same minimum number of change patterns needed to
construct the solution.


\textbf{Response Variables.} To operationalize cognitive complexity of change
patterns usage we consider, (1) accuracy, i.e., how close the subject's solution
is to the reference model, (2) efficiency, i.e., how many detours it takes them
to reach the solution, (3) speed, i.e., how fast they create the solution, and
(4) the required mental effort. To measure accuracy we consider \textit{product
deviations}, i.e., discrepancies between the modeler's solution and the reference
model. For example, a process model which contains two product deviations,
requires the application of two change patterns to transform that model into the
reference model. To operationalize efficiency, we consider \textit{process
deviations} measuring the modeler's detours until coming up with the solution.
Process deviations are calculated as difference between the number of applied
change patterns to reach the solution and the minimum number of change patterns
required for the task. Finally, we consider the \textit{time} needed to
accomplish the task (i.e., speed) as well as the required \textit{mental effort},
measured using a questionnaire after the task~\cite{ZPRR+12}.

\textbf{Hypotheses.} This leads us to the following null hypotheses.

\vspace{-0.2cm}
\begin{itemize}
\item $H_{1,0}$: High nesting depth does not lead to significantly more product
deviations when compared to low nesting depth.
\item $H_{2,0}$: High nesting depth does not lead to significantly more process
deviations when compared to low nesting depth.
\item $H_{3,0}$: High nesting depth does not require significantly more time
when compared to low nesting depth.
\item $H_{4,0}$: High nesting depth does not impose a significantly higher
mental effort when compared to low nesting depth.
\end{itemize}  
\vspace{-0.2cm}

\textbf{Instrumentation. } For data collection
Cheetah Experimental Platform \cite{PZW10} is used, logging given answers
(e.g., demographic data), the time to accomplish the tasks, and all model
interactions to obtain process deviations.

\textbf{Experimental Design. }  The experiment is conducted as
balanced single factor experiment with repeated measurements.  Prior to the
experiment a familiarization phase takes place (i.e., subjects are 
trained using change patterns). Subjects are then randomly assigned to two
groups of equal size, subsequently referred to as G1 and G2. To
provide a balanced experiment with repeated measurements, the overall procedure
consists of two runs. In the first run G1 applies factor level \textit{low
nesting depth}, G2 factor level \textit{high nesting depth}. In the second
run, factor levels are switched and G1 applies factor level \textit{high
nesting depth}, G2 factor level \textit{low nesting depth} to the same
object. Choosing such a cross design is an additional measure to counter
potential learning effects.

\vspace{-0.4cm}
\section{Summary} 
\vspace{-0.1cm}

While the process of creating process models using change
primitives has caused some interest in recent years \cite{Sof+12,PZW+12,CVR+12,PSZ+12}, our
understanding of the process of creating process models using change patterns is
limited. This paper proposes a research design to investigate the PPM using
change patterns in more detail. In particular, the impact of nesting depth on
the cognitive complexity of creating models is examined. Results of the
experiment will provide a better understanding of the PPM using change patterns
and help to understand how to design tool-support for
change patterns based modeling.
\vspace{-0.4cm}

\bibliography{literature}
\bibliographystyle{splncs}
\end{document}